\documentclass[doublecol]{epl2}

\usepackage{ifthen}
\usepackage{ifpdf}
\usepackage{color}

\ifpdf
\usepackage{graphicx}
\usepackage{epstopdf}
\else
\usepackage{graphicx}
\usepackage{epsfig}
\fi

\usepackage{latexsym}
\usepackage{amsmath}
\usepackage{amssymb}
\usepackage{bm}
\usepackage{wasysym}

\newcommand{\eexp}{\mbox{e}^}

\newcommand{\tbox}[1]{\mbox{\tiny #1}}
 
\newcommand{\mylabel}[1]{\label{#1}} 
\newcommand{\beq}{\begin{eqnarray}}
\newcommand{\eeq}{\end{eqnarray}} 
\newcommand{\be}[1]{\begin{eqnarray}\ifthenelse{#1=-1}{\nonumber}{\ifthenelse{#1=0}{}{\mylabel{e#1}}}}
\newcommand{\ee}{\end{eqnarray}} 

\newcommand{\Eq}[1]{\textcolor{blue}{Eq.\!\!~(\ref{#1})}} 
\newcommand{\Fig}[1]{\textcolor{blue}{Fig.}\!\!~\ref{#1}} 

\title{Non-equilibrium steady state of sparse systems}
\shorttitle{NESS of sparse systems}

\author{Daniel Hurowitz and Doron Cohen}

\institute{
Department of Physics, Ben-Gurion University of the Negev, P.O.B. 653, Beer-Sheva 84105, Israel
}

\abstract{
A resistor-network picture of transitions is appropriate  
for the study of energy absorption by weakly chaotic or weakly 
interacting driven systems. Such ''sparse" systems
reach a novel non-equilibrium steady state (NESS) once coupled to a bath. 
In the stochastic case there is an analogy to the physics 
of percolating glassy systems, and an extension 
of the fluctuation-dissipation phenomenology is proposed.  
In the mesoscopic case the quantum NESS might 
differ enormously from the stochastic NESS, 
with saturation temperature determined by the sparsity.
A toy model where the sparsity of the system is modeled 
using a log-normal random ensemble is analyzed. 
}

\pacs{}{}

\begin{document}
\maketitle

\begin{figure}[h!]
\centering

(a) \hspace*{0.6\hsize} \\
\includegraphics[clip, width=0.8\hsize]{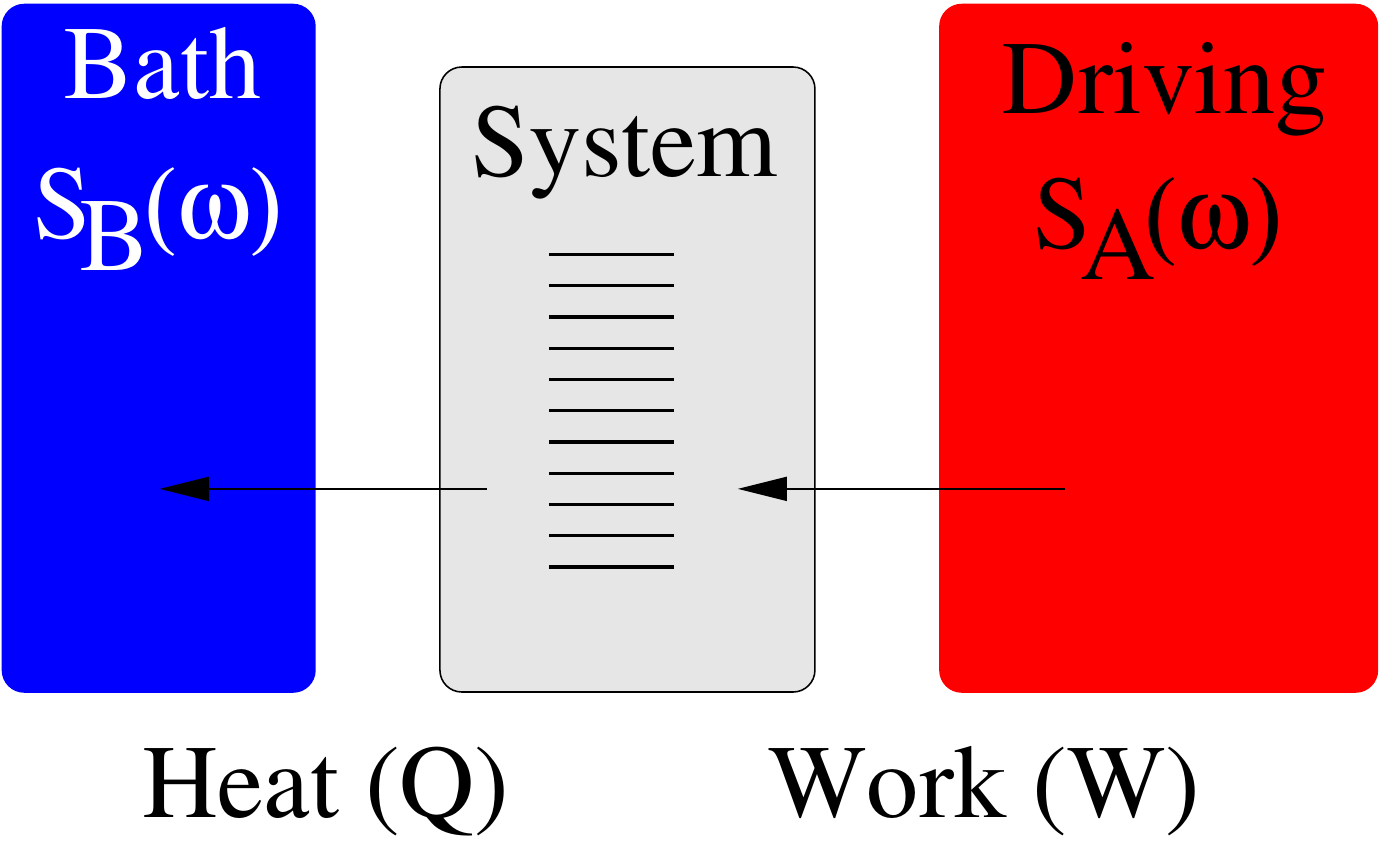}

\vspace*{5mm}

(b) \hspace*{0.6\hsize} \\
\includegraphics[clip, width=0.8\hsize]{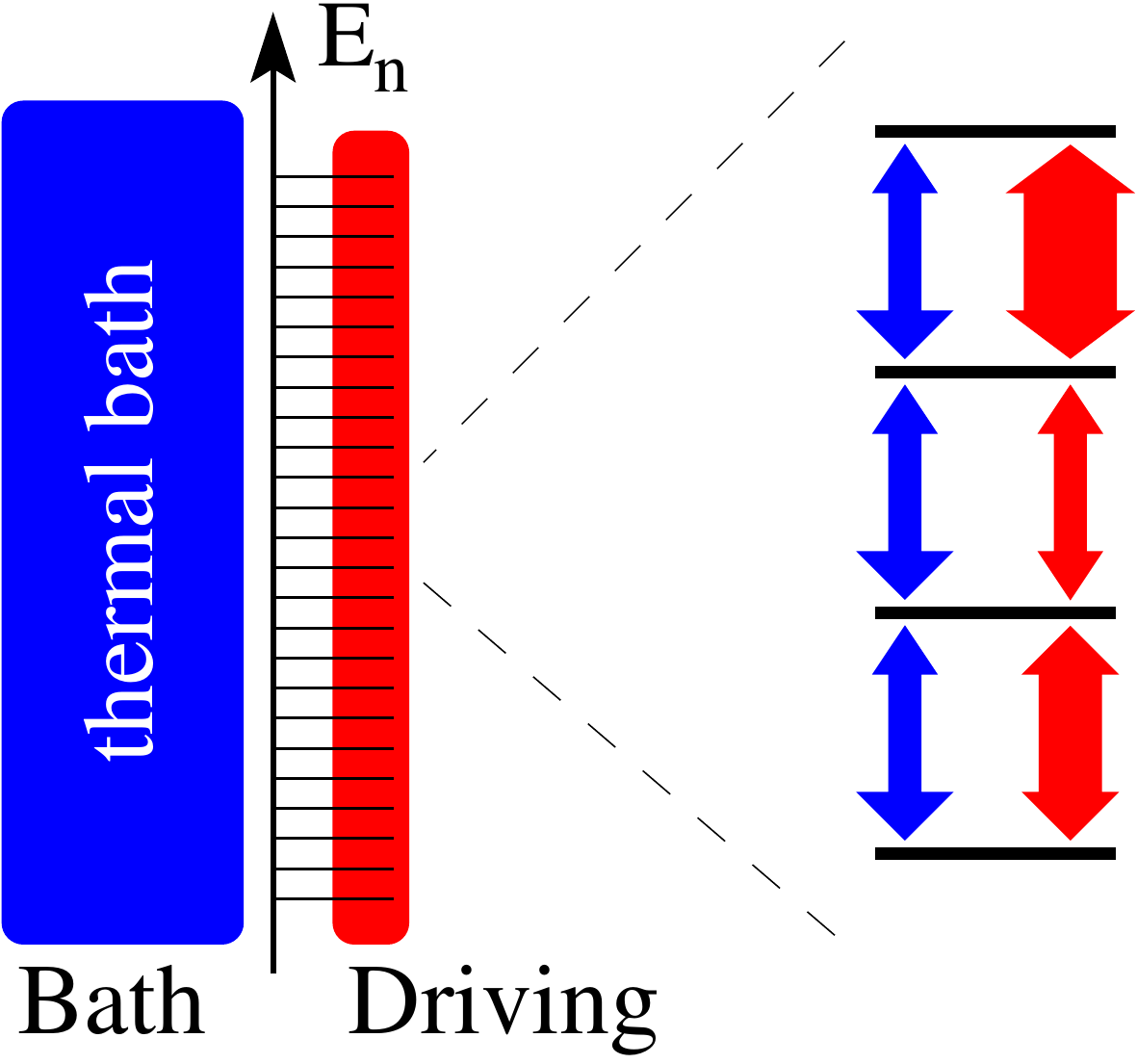}

\vspace*{5mm}

\caption{
(a) Block diagram of the model system.
Both the driving source (A) and the bath (B) induce 
fluctuations that drive transitions. 
These fluctuations are characterized by a 
power spectrum $\tilde{S}(\omega)$. 
The driving source can be formally regarded 
as a bath that has an infinite temperature ${T_A=\infty}$, 
such that $\tilde{S}_A(-\omega)=\tilde{S}_A(\omega)$.
In contrast to that $T_B$ is finite, 
and accordingly  $\tilde{S}_B(-\omega)/\tilde{S}_B(\omega)=\exp(\omega/T_B)$. 
\ \ \ 
(b) Illustration of the Chain model.
In the absence of a bath the driving induces transitions (red arrows) between 
levels $E_n$ of a closed system, leading to diffusion 
in energy space and hence heating. 
The diffusion coefficient $D$ can be calculated 
using a resistor network analogy. 
A NESS is reached due to the presence of a heat bath (blue arrows) 
that favors downward transitions.  
}

\label{f1}
\end{figure}

The study of systems with non-equilibrium steady state (NESS) 
has become active in recent years \cite{ness1,ness2,ness3,mixed1,mixed2,mixed3},  
involving various generalizations of linear response theory (LRT) 
and of the associated fluctuation-dissipation relation (FDR) 
\cite{EGD1,crooks,chris,irrev1,glass,violation,langevin1}. 
The paradigm for NESS (\Fig{f1}a) is a system that is coupled to 
two equilibrated reservoirs, "A" and "B", 
which are characterized by different temperatures $T_A$ and $T_B$. 
Hence the NESS of the system is not canonical,
and it cannot be characterized by a well-defined equilibrium temperature.
A particular case of special interest is obtained 
if one reservoir (call it "A") is replaced by a stationary 
driving source, while the relaxation is provided by 
a bath (call it "B") that has some finite temperature~$T_B$. 
This is still the same paradigm because formally the 
driving source "A" can be regarded as a bath that has 
an infinite temperature $T_A=\infty$.

{\bf Stochastic modeling.--} 
The simplest modeling of NESS is obtained by considering a system 
that has energy levels $\{E_{n}\}$ with transition rates
\be{1}
\mathcal{W}_{nm} = w^{\varepsilon}_{nm} + \frac{2w_{nm}^{\beta}}{1+\eexp{(E_n{-}E_m)/T_B}} 
\ee 
where $w^{\varepsilon}_{nm}$ and $w_{nm}^{\beta}$ 
are the elements of symmetric matrices. The first term 
describes the transitions that are induced by the $T_A{=}\infty$ driving source.
The second term describes the bath induced transitions, 
with ratio $\eexp{(E_n{-}E_m)/T_B}$ of $n\Leftrightarrow m$ transitions,  
as required by detailed balance considerations.
The dynamics of the population probabilities $\bm{p}=\{p_n\}$ 
is described by a rate equation 
\be{222}
\frac{dp_n}{dt} = \sum_m \left[\mathcal{W}_{nm}p_m - \mathcal{W}_{mn}p_n\right]
\eeq 
This equation can be written schematically 
as ${\dot{\bm{p}}=\mathcal{W}\bm{p}}$, see remark~\cite{a}. 
The steady state is determined from the matrix equation ${\mathcal{W}\bm{p}=0}$. 
In the presence of driving the detailed balance is disturbed leading in general 
to a non-canonical NESS.

{\bf Sparse systems .--} 
In recent studies \cite{kbr,slr,kbw} our interest 
was focused on a class of driven systems 
for which the matrix $\{w_{nm}^{\varepsilon}\}$ is {\em sparse}.
By this we mean that the transition rates are 
characterized by a log wide (say log normal) distribution.
In other words, the majority of 
elements are small, while the large elements constitute 
a small fraction, ${s\ll1}$. 
A system of current experimental interest is described in \cite{kbw}: 
an optical billiards with vibrating walls, 
where the energy absorption rate  {(EAR)} 
of the cold atoms is affected by the sparsity 
of the perturbation matrix, which is controlled either 
by the degree of chaoticity or by the strength of the inter-atomic interactions.        
Yet a simpler example concerns the absorption of low frequency irradiation
by small metallic grains~\cite{slr}, where the transitions are predominately 
between neighboring energy levels, and the  {sparsity} is determined by 
the level spacing statistics.  

{\bf Beyond LRT .--} 
In the absence of a bath 
the rate equation ${\dot{\bm{p}}=\mathcal{W}\bm{p}}$ generates 
{\em diffusion} in energy space. In this context it 
is useful to picture the levels~$n$ as ``sites" in space, 
or as the ``nodes" of a network,  
and the $w_{nm}^{\varepsilon}$ as ``connectors" (\Fig{f1}b). 
The calculation of the diffusion coefficient~$D$     
is exactly as the calculation of electrical conductivity.
For example, when connecting $N$ nodes in {\em series} 
(as in the Chain model to be discussed later), 
the ``conductivity" is ${D=[(1/N)\sum_n (1/w_n)]^{-1}}$.
The adaptation of the resistor network picture to the 
calculation of~$D$ is termed semi-linear response theory (SLRT) 
because~$D$ is a semi-linear function of the couplings~$w_{nm}$.
This means that $D[c\bm{w}]=cD[\bm{w}]$, but there is 
no additivity, $D[\bm{w}+\bm{w'}]\neq D[\bm{w}]+D[\bm{w'}]$. 
Due to the sparsity, the result is very similar to that of 
percolating or disordered resistor-networks \cite{miller}.

{\bf Outline.--} 
Considering the coupling of a ``sparse" system to a bath,  
our expectation is to have, 
as the driving becomes stronger, 
a crossover from an LRT canonical-like NESS 
to a novel non-canonical NESS, 
with the possibility of remarkable quantum-mechanical fingerprints. 
Specifically our objective is to calculate 
the {\em energy absorption rate} {(EAR)} of a driven ``sparse" system, 
taking into account the presence of a surrounding environment.  
Below we  
{\bf (i)} introduce the FDR phenomenology of calculating the EAR, 
which requires a notion of effective NESS temperature; 
{\bf (ii)} demonstrate this phenomenology for the simplest toy model, 
obtaining explicit expressions for both $T$ and $D$; 
and {\bf (iii)} discuss the quantum case, highlighting the existence 
of a saturation temperature $T_{\infty}$ that is determined by the sparsity.

{\bf FDR phenomenology.--} 
In the {\em standard} textbook presentation 
it is assumed that the system reaches a canonical-like state 
with a well-defined temperature~$ {T_{\tbox{sys}}}$. 
The driving induces diffusion with coefficient~${D}$ 
in energy space \cite{ott1,jar,wilk,robbins,frc}.  
From ${\dot{E}=\sum_n E_n \dot{p}_n}$, 
substituting ${\dot{\bm{p}}=\mathcal{W}\bm{p}}$,
it follows (see appendix) that the  {EAR} is 
\be{2}
\dot{\mathsf{W}} 
\ = \ 
\sum_{n,m} (E_n-E_m) w^{\varepsilon}_{nm}p_{m} 
\ = \
\frac{D}{ {T_{\tbox{sys}}}}
\eeq
with 
\be{333}
D\mbox{\tiny [LRT]} = \frac{1}{2} \overline{\sum_{n} w_{nm}^{\varepsilon} (E_n{-}E_m)^2} 
\eeq 
where the overline indicates canonical averaging over the initial state. 
In complete analogy it is straightforward to show (see appendix), 
that the rate of cooling due to the interaction with the bath can be written as 
\be{3}
\dot{\mathsf{Q}}
\ = \ 
- \sum_{n,m} (E_n-E_m)w^{\beta}_{nm}p_{m} 
\ = \ 
\frac{D_B}{T_B}-\frac{D_B}{ {T_{\tbox{sys}}}} 
\eeq
where the first term is due to the imbalance 
of upward and downward transition rates, 
while the second term is due to the non-uniformity  
of the probability distribution in energy space.  
The net rate of energy increase is $\dot{E}=\dot{\mathsf{W}}-\dot{\mathsf{Q}}$.
At steady state ${\dot{\mathsf{W}}=\dot{\mathsf{Q}}}$, 
so a phenomenological determination 
of the steady state temperature~$ {T_{\tbox{sys}}}$ is possible.

The essence of the above FDR phenomenology is the same 
as in Einstein's relation: the dissipation is related to 
the diffusion (in LRT the latter is determined by the 
fluctuations, e.g. the velocity-velocity correlation 
in Einstein's relation, hence the terminology). 
Our purpose below is to {\em generalize} the FDR.
We emphasize in advance two issues:
{\bf (a)} The NESS might be non-canonical, 
so we have to define its effective temperature;
{\bf (b)} The diffusion coefficient is not 
necessarily determined by LRT. 

{\bf The Chain model.--} 
It is best to clarify the determination of~$ {T_{\tbox{sys}}}$ and~$D$ by considering 
the simplest example of a ``chain" with nearest-neighbor transitions only.
With simplified indexing, \Eq{e1} for ${(n{-}1) \Leftrightarrow n}$ transitions 
is written as ${w_n+2w^{\beta}/(1+\eexp{\pm\Delta_0/T_B})}$,   
where ${\Delta_0}$ is the level spacing, and ${n=1,2,\cdots}$.
In contrast to $w^{\beta}$, which is the same for all transitions, 
the rates $w_n$ are characterized by a logarithmically~wide distribution.

In the numerical example we have $N{=}25$ levels, 
with equal level spacing $\Delta_0{=}1$.
The bath temperature is $T_B{=}10$.
The bath induced transition rates 
are taken as ${w^{\beta}=0.1}$.   
The driving induced transition rates are log normally distributed. 
Namely, $w_n=\exp(x_n)$, where the $x_n$ have a Gaussian distribution 
with average $\mu$ and dispersion $\sigma$ that are determined 
such that the driving intensity is $\overline {w_n}  = \varepsilon^2$,  
and the sparsity  \cite{kbw} is ${s=\exp(-\sigma^2)}$.
The value ${s\sim1}$ means that all the elements 
are comparable and well represented by their average. 
Sparsity means ${s\ll1}$, for which the median differs 
by orders of magnitude from the algebraic average.

\begin{figure}[h!t]

\includegraphics[width=0.7\hsize]{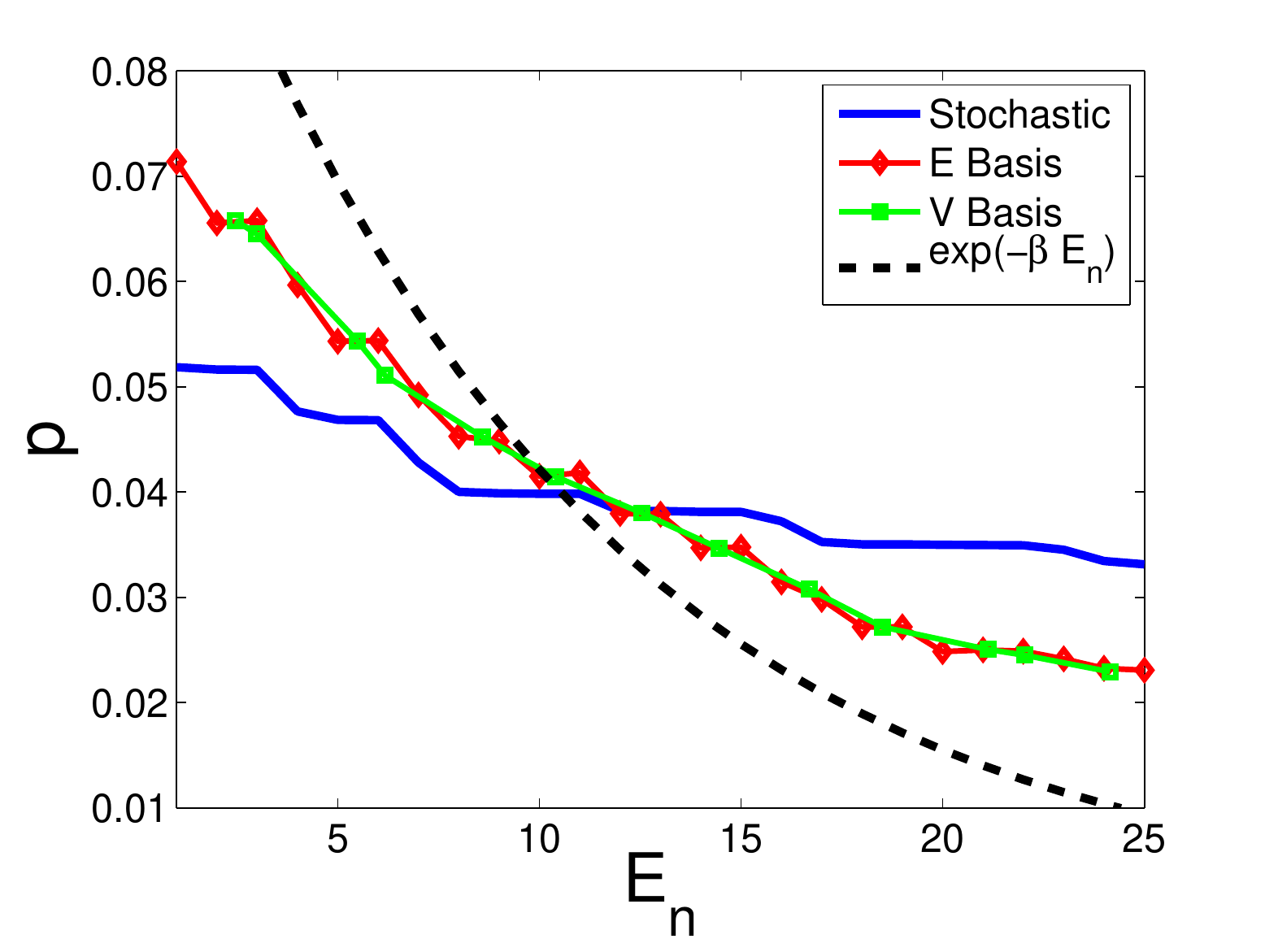}
\centering

\caption{
The NESS occupation probabilities $p_n$ are plotted vs $E_n$ 
in the stochastic (blue) and quantum (red) cases.  
In the latter (quantum) case we plot also the occupation probabilities $p_r$ 
of the $V$ eigenstates versus $\langle E\rangle_r$.  
The sparsity is $s = 10^{-5}$, and $\varepsilon=9.3$. 
We observe that the effective temperature predicted by the quantum master equation 
is lower compared with the stochastic approximation.
}

\label{fP}
\end{figure}

\begin{figure}[h!t]
\centering

\includegraphics[width=0.7\hsize]{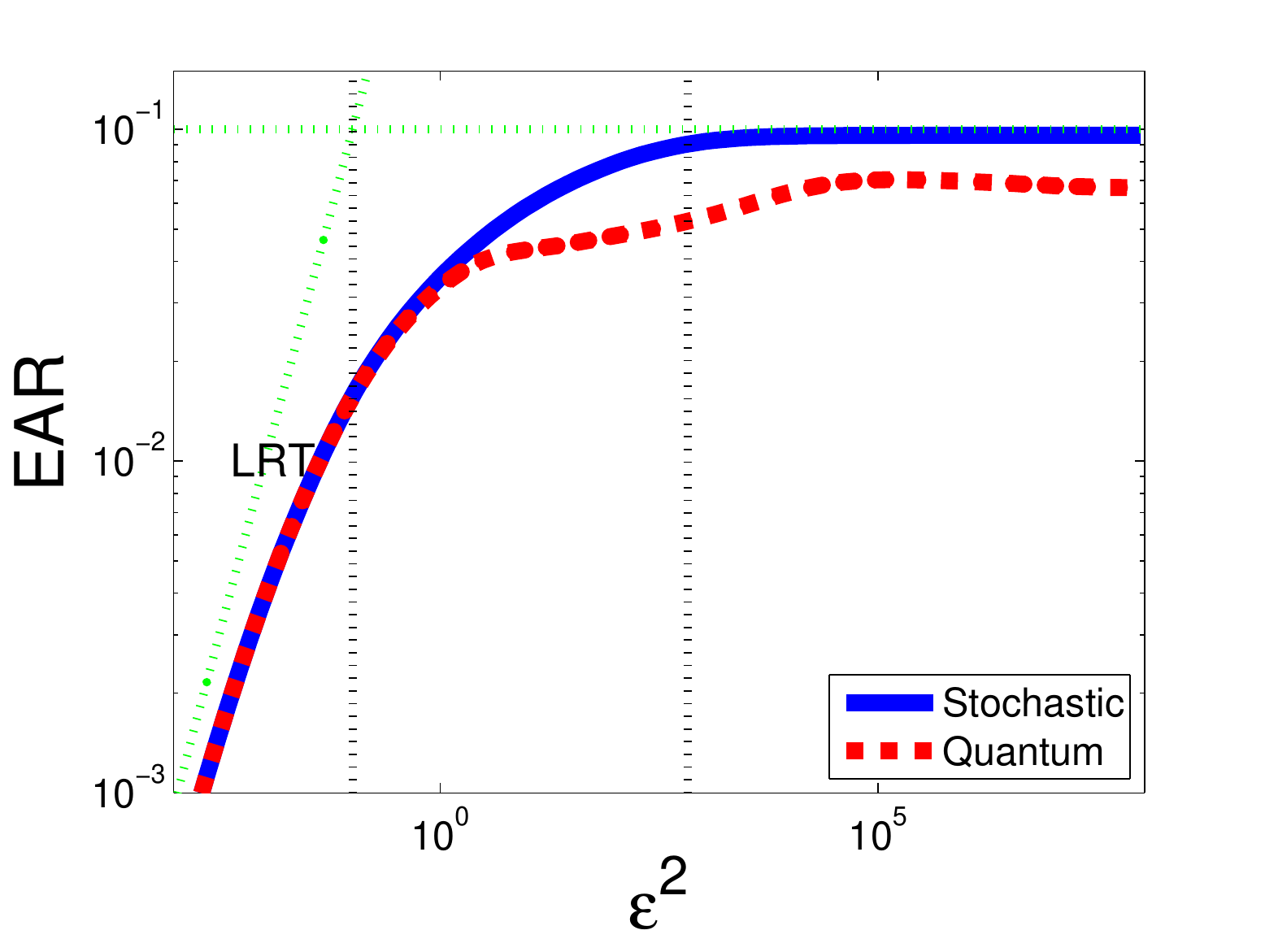}

\caption{
The  {EAR} $\dot{\mathsf{W}}$ 
for the stochastic (solid blue) 
and for the quantum (dashed red) NESS. 
The sparsity is ${s=10^{-5}}$. 
The vertical lines are plotted at values of $\varepsilon$ 
for which the stochastic picture predicts a crossover: 
i.e. the $\varepsilon$ values for which $\overline{w_n}$ 
and $\overline{[1/w_n]^{-1}}$ equal $w^{\beta}$.  
}

\label{fW}
\end{figure}

From the NESS equation ${\mathcal{W}\bm{p}=0}$  
we determine the occupation probabilities~$p_n$ as in \Fig{fP}, 
and then calculate the  {EAR} via either \Eq{e2} or \Eq{e3} as in \Fig{fW}. 
In the absence of driving, the steady state is canonical 
with a well-defined temperature $T_B$. In the presence 
of driving, the state is generally not canonical. 
Consequently, we can formally associate a different microscopic temperature $T_{nm}$ 
for each pair of coupled levels via the defining equation 
\beq
\frac{p_n}{p_m} \ \ = \ \ \exp\left(-\frac{E_n-E_m}{T_{nm}}\right)
\eeq    
For the Chain model we use the simpler indexing $T_n$,  
and the NESS equation ${\mathcal{W}\bm{p}=0}$ 
takes the form $p_n/p_{n{-}1}=\mathcal{W}_{n,n{-}1}/\mathcal{W}_{n{-}1,n}$.    
Assuming $\Delta_0 \ll T_B$ we deduce that   
the microscopic temperature of the n${th}$ transition 
is given by the expression 
\be{777}
T_n = \left[\frac{w_{n}+w^{\beta}}{w^{\beta}}\right]T_B
\eeq

\begin{figure}[h!t]
\centering

\includegraphics[width=0.7\hsize]{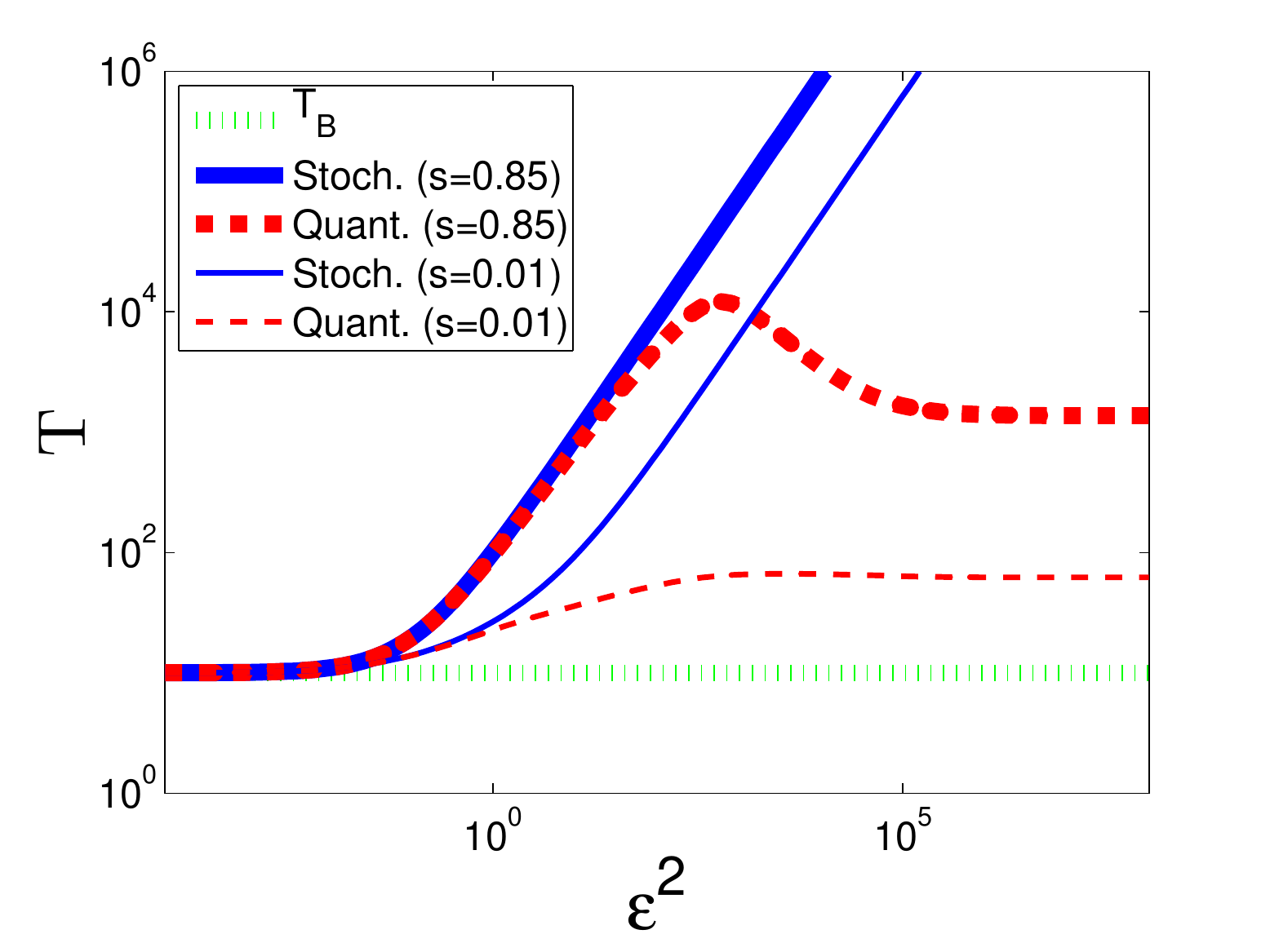}

\caption{
The effective NESS temperature $ {T_{\tbox{sys}}}$ 
versus the driving intensity.
Solid (blue) lines are for the stochastic NESS, 
while dashed (red) lines are for the quantum NESS. 
The dotted (green) line represents the temperature~$T_B$ of the bath.
}

\label{fT}
\end{figure}

{\bf Effective NESS temperature.--} 
We define the NESS temperature~$ {T_{\tbox{sys}}}$ such that the 
phenomenological FDR \Eq{e3} still holds.
For the Chain model the bath induced diffusion is $D_B=w^{\beta}\Delta_0^2$,  
and it is straightforward to show (see appendix) that~$ {T_{\tbox{sys}}}$ 
should be defined as the harmonic average over $T_n$, i.e. 
\be{50}
 {T_{\tbox{sys}}} =  \left[\overline{\left(\frac{1}{T_n}\right)}\right]^{-1}
= \left[\overline{\left(\frac{w^{\beta}}{w^{\beta}+w_n}\right)}\right]^{-1} T_B
\eeq
For numerical results see (\Fig{fT}-5). As the driving becomes stronger, 
the temperature becomes higher, with the asymptotic 
behavior ${ {T_{\tbox{sys}}} \propto \varepsilon^2}$ as implied by \Eq{e50}. 

{\bf LRT-SLRT crossover.--} 
We continue with the stochastic Chain model and find from \Eq{e2},
substituting \Eq{e777} in \Eq{e20} of the appendix,  
that the  {EAR} is given by the expression 
\be{242}
\dot{\mathsf{W}} \ \ = \ \ \left[\overline{\left(\frac{w_n}{w^{\beta}+w_n}\right)}\right] \frac{D_B}{T_B}
 \ \ \equiv \ \  \frac{D}{ {T_{\tbox{sys}}}}
\eeq
We define the driving induced  {diffusion}~$D$ such that the 
phenomenological FDR still holds:
\be{243}
D=    
\left[\overline{\left(\frac{w_n}{w^{\beta}+w_n}\right)}\right] 
\left[\overline{\left(\frac{1}{w^{\beta}+w_n}\right)}\right]^{-1}
\Delta_0^2
\eeq
In the limit of very weak and very strong driving we respectively get 
\be{4}
D\mbox{\tiny [LRT]} &=&\overline{w_n}\Delta_0^2 
\\ \label{e5}
D\mbox{\tiny [SLRT]} &=&[\overline{1/w_n}]^{-1}\Delta_0^2 
\eeq
The LRT result could be obtained from the traditional Kubo formalism,
while the SLRT prediction reflects a network that consists 
of connectors that are connected in series. 
Note that if all the $w_n$ are comparable,  
then ${D \approx D\mbox{\tiny [LRT]} \approx D\mbox{\tiny [SLRT]}}$. 
But if the $w_n$ have a log-wide distribution,  
the agreement with LRT is achieved only if 
the $w_n$ are all much smaller than $w^{\beta}$. 
For strong driving, both $D$ and $ {T_{\tbox{sys}}}$ 
are $\propto [\overline{1/w_n}]^{-1}$, 
and hence  ${\propto \varepsilon^2}$, 
as expected from the SLRT resistor-network phenomenology. 
In this limit their ratio approaches the bath limited value
%
\be{1333}
\dot{\mathsf{W}}_{\infty} \ \ = \ \ \frac{D_B}{T_B} 
\eeq
which is implied by \Eq{e3}.

{\bf Quantum modeling.--} 
It should be clear that SLRT applies whenever the transport  
is modeled using a resistor network. Thus it might have 
applications, e.g., in statistical mechanics and biophysics.
But the original motivation for SLRT came from mesoscopics, 
where the quantum nature of reality cannot be ignored.
In this context the Hamiltonian contains a driving term $f(t)V$, 
and the transition rates $w_{nm}^{\varepsilon} \propto |V_{nm}|^2$ 
between levels are determined by the Fermi-Golden-Rule (FGR). 
For weakly chaotic or weakly interacting systems $V_{nm}$ 
is typically sparse and textured. 
Assuming that $f(t)$ has zero average, and correlation 
function ${\langle f(t)f(t') \rangle = \varepsilon^2 \delta(t-t')}$, 
the rate equation \Eq{e222} is replaced by the quantum Master equation 
\be{10}
\frac{d\rho}{dt}=-i[\mathcal{H}_0,\rho]-\frac{\varepsilon^2}{2}[V,[V,\rho]] + \bm{\mathcal{W}}^{\beta}\rho
\eeq
where the second and third terms correspond to the driving source 
and to the bath as in \Eq{e1}. This equation is of 
the form ${\dot{\rho} =\bm{\mathcal{W}}\rho}$. 
It is identical with ${\dot{\bm{p}}=\mathcal{W}\bm{p}}$ of \Eq{e222}
if the off-diagonal terms are ignored, where 
\beq
w_{nm}^{\varepsilon} \ \ = \ \ \varepsilon^2 |V_{nm}|^2
\eeq
Note that $\bm{\mathcal{W}}^{\beta}$ induces dephasing 
of the off-diagonal elements, which we assume to be minimal 
in the ``quantum" case: see technical details below.

Some technical details with regard to \Eq{e10} are in order 
(can be skipped in first reading).   
In the energy basis, $\mathcal{H}_0$ is a diagonal matrix with energy levels~$E_n$.  
The state of the system is represented by the probability 
matrix, which can be rewritten as a column vector ${\rho \mapsto (p_n;\rho_{\nu\mu})}$,  
composed of the diagonal {\em probabilities} and the off-diagonal {\em coherences}. 
Consequently, the master equation takes the form ${\dot{\rho} =\bm{\mathcal{W}}\rho}$, 
with the super operator
\be{11}
\bm{\mathcal{W}} 
=
\left( \begin{array}{cc}
\mathcal{W} & \Lambda^{\dag}  \\ 
\Lambda & \mathcal{W}^{\perp} \\ 
\end{array} \right)
\eeq
The matrix $\mathcal{W}$ is given by \Eq{e1} 
with $w_{nm}^{\varepsilon}=\varepsilon^2|V_{nm}|^2$. 
The definition of $\Lambda$ is implied by the 
second term in \Eq{e10}. In particular we note that 
\be{14}
\mathcal{W}^{\perp}_{\nu\mu,\nu\mu} &=&  i\Delta_{\nu\mu} -\gamma_{\nu\mu} -\gamma_{\beta}
\\ \label{e15}
\Lambda_{n,\nu\mu} &=& \varepsilon^2 V_{n\nu}V_{\mu n},  \ \ \ \ \mbox{for $\nu,\mu \ne n$}
\eeq    
We use the common notations $\Delta_{\nu\mu}=E_{\nu}{-}E_{\mu}$, 
and $\gamma_{\nu\mu}=(\varepsilon^2/2)[(V^2)_{\nu\nu}{+}(V^2)_{\mu\mu}]$. 
For simplicity, we assume that the bath induced 
dephasing  {${\gamma_{\beta} =  w^{\beta} + \gamma_{\varphi}}$} 
is the same for all the coherences. 

 {In the absence of driving} \Eq{e10} is the well known Pauli 
master equation leading to canonical equilibrium.  
 {In the numerics} we assume in the ``quantum" case minimal dephasing,   
which means ${\gamma_{\varphi}=0}$.  {More generally} we may have 
an extra ``pure dephasing" effect which is represented by ${\gamma_{\varphi}>0}$
as in the familiar NMR context. 

The stochastic (FGR) picture applies if the coherences 
become negligible. By inspection, it might happen   
for two reasons.  {One possibility is that there is 
strong pure dephasing effect} (${\gamma_{\varphi}\gg w^{\varepsilon}}$) 
that suppresses the coherences. Then one simply recovers  
the stochastic rate equation for the occupation probabilities.   
The second possibility is to have $w^{\varepsilon}$ much smaller 
compared with the level spacing~$\Delta_0$. 
The latter is the traditional assumption in atomic physics, 
and can be regarded as ``microscopic circumstances". 
But in ``mesoscopic circumstances" $\Delta_0$ might be small, 
and the validity of FGR is not guaranteed.

{\bf Quantum NESS.--} 
The NESS is obtained  by solving the equation ${\bm{\mathcal{W}}\rho=0}$. 
Looking at \Fig{fT}-5 we see that in the ``quantum" case there 
is a saturation temperature ${T_B<T_{\infty}<\infty}$ that depends on the sparsity~$s$.
Consequently there is a premature saturation of the  {EAR} at 
%
\be{1999}
\dot{\mathsf{W}}_{\infty} \ = \ \frac{D_B}{T_B}-\frac{D_B}{T_{\infty}}
\eeq
as seen in \Fig{fW}.  Note \cite{b}.
Only in the non-sparse limit do we recover quantum-to-classical correspondence.

\begin{figure}[h!t]
\centering

\includegraphics[width=0.7\hsize]{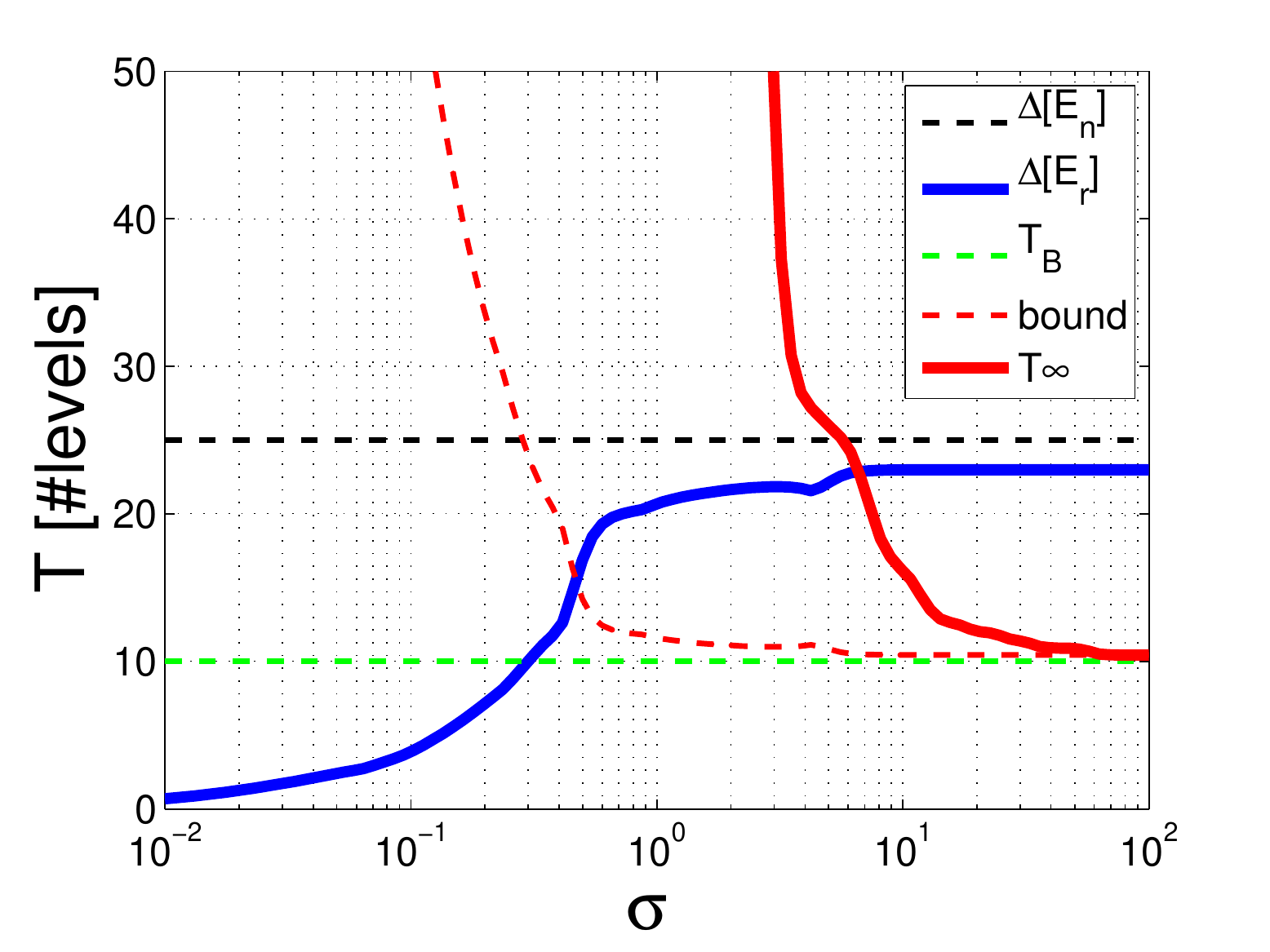}
  
\caption{ 
The dependence of $T_{\infty}$ on the width $\sigma$ of the log-normal distribution.
Note that the sparsity is ${s=\exp(-\sigma^2)}$. 
We confirm that $T_{\infty}$ is bounded from below  
by ${[\Delta(E_n)/\Delta(E_r)] T_B}$ (dashed red line),  
and tends to ${T_B}$ in the sparse limit. 
Here $\Delta(E_n)=25$ is the width of energy window in 
this numerical test, while $\Delta(E_r)$ is the length of 
the interval that contains the energies $\langle E \rangle_r$.
}

\label{TvsSigma}
\end{figure}

One would like to understand how~$T_{\infty}$ emerges.
For this purpose we regard \Eq{e10} for $\rho_{nn'}$ 
as a Fokker-Planck equation, where~$n$ is the momentum. 
In the absence of sparsity, $V_{nn'}$ can be interpreted 
as the matrix representation of the position coordinate, 
and its eigenstates are extended in~$n$. 
But if $s\ll1$, then $V_{nn'}$ is like off-diagonal disorder, 
and its eigenstates~$r$ become localized in~$n$, 
with energies $\langle E \rangle_{r}$ that are no longer identical. 
If the driving is very strong, the master equation implies that 
the NESS becomes a mixture of the eigenstates~$r$ 
with weights 
\be{0}
p_r \ \propto \ \exp\left[- \frac{\langle E \rangle_{r}}{T_{\tbox{mix}}}\right]
\eeq
In the sparse limit the~$n$ and the~$r$ bases essentially coincide, 
and accordingly ${T_{\infty} \sim T_{mix} \sim T_B}$.
This is confirmed numerically in \Fig{TvsSigma}.

In order to further illuminate the variation of the 
saturation temperature as a function of ${s=\exp(-\sigma^2)}$, 
we have added in \Fig{TvsSigma} a lower bound estimate 
for $T_{\infty}$ which is deduced as follows \cite{c}: 
Considering an energy range $\Delta(E_n)$ of interest, 
one observes that as $\sigma$ is decreased, 
the $\langle E \rangle_{r}$ occupy a {\em smaller} range $\Delta(E_r)$, 
and eventually in the non-sparse limit they are quasi-degenerate.
The~$p_r$ distribution is characterized by some ${T_{mix} \geq T_B}$.
Consequently the stretched distribution $p_n$ is characterized by 
a higher temperature ${T_{\infty} \geq [\Delta(E_n)/\Delta(E_r)]T_{B}}$.
Hence the saturation temperature increases for smaller $\sigma$, 
and diverges in the non-sparse limit.

 {{\bf Experimental Signature.--}} 
The {\em thermodynamic} definition of temperature which we have adopted 
via \Eq{e3} is related to the heat flow between two bodies: 
it is the same concept which is reflected in the phrasing of 
``the zeroth law of thermodynamics" and in the associated definition 
of {\em empirical} temperature. It should be contrasted 
with an optional {\em statistical mechanics} perspective that 
relates the temperature to the fluctuations, as in the paradigmatic 
Brownian motion studies of Einstein. 
Once $T_{\tbox{sys}}$ of \Eq{e3} is determined through 
a heat flow measurement, it is possible to determine experimentally 
also the $D$ in \Eq{e2} through an EAR measurement. 
Then it is possible to explore the LRT to SLRT crossover 
of this absorption coefficient \Eq{e4}-\Eq{e5}. This crossover is reflected in having a 
wide range (many decades) over which the EAR varies as in \Fig{fW}.
It is important to realize that the unlimited increase 
of temperature in the classical case does not have by itself 
a reflection in the EAR, because classically the EAR always 
saturates to a {\em bath~limited~value} as determined by \Eq{e1333}. 
In contrast to that, quantum mechanically there is a premature 
saturation, as expected from \Eq{e1999}, reflecting the sparsity 
of the system. This effect might be useful for the detection of 
a classical to quantum transition with non-equilibrium measurements.

{\bf Discussion.--} 
The ``sparse" NESS resembles that of a glassy system~\cite{glass}: 
In both cases, one has to distinguish between ``microscopic" 
and ``macroscopic" time scales, rates, and temperatures. 
\Eq{e2} and \Eq{e3} establish a diffusion-dissipation 
relation involving a ``macroscopic" temperature~$T$ that might be much 
lower compared with the microscopic temperatures~$T_{n}$. 
The diffusion is driven by the fluctuations of the sources, 
but it is not the LRT (Kubo) formula which should be used in order to determine~$D$, 
but rather a resistor network SLRT calculation. 

A closely related chain of works regarding NESS, concerns mixed phase-space 
of periodically driven systems~\cite{mixed1,mixed2,mixed3}, 
where the problem is reduced to the study 
of a stochastic rate equation. Our work differs in three respects: 
(1)~The sparsity may arise even for quantized chaotic 
non-mixed systems, implying a glassy type of NESS;
(2)~We assume a stationary driving source 
instead of a strictly periodic driving;
(3)~The master equation approach has allowed us to consider 
novel mesoscopic circumstances in which the quantum NESS differs 
enormously from the stochastic prediction.
     
The influence of quantum coherence on the NESS is remarkable.  
Due to the localization of the eigenstates 
in energy space, we found that for strong driving 
the temperature saturates to a {\em finite} value  
that reflects the sparsity of system. This should be 
contrasted with the traditional prediction of unbounded temperature. 

{\bf Appendix.--} 
For the convenience of the reader we concentrate here the technical 
details that concern the derivation of the FDR phenomenology.
The energy of the system is ${E=\sum_n p_n E_n}$, 
and its rate of change is ${\dot{E}=\sum_n E_n \dot{p}_n}$.
The equation for $\dot{p}_n$ includes a driving source  
term and a bath term. Accordingly we 
write $\dot{E}=\dot{\mathsf{W}}-\dot{\mathsf{Q}}$, where the 
two terms are interpreted as the rate of {\em heating} due 
to the driving  {(equals the EAR)}, 
and the rate of {\em cooling} due to the bath 
 {(in steady state equals the EAR too)}. 
From the master equation it follows that the 
expression for $\dot{\mathsf{Q}}$, both in the stochastic 
and in the quantum case, is given by the first equality 
of \Eq{e3}.  This expression can be written as the sum 
of a term that originates from the asymmetry of $w^{\beta}_{nm}$,  
and a term that originates from the non-uniformity of the $p_{n}$. 
Defining ${\bar{p}_{nm} = (p_n{+}p_m)/2}$ we get
\beq
w^{\beta}_{nm}-w^{\beta}_{mn} 
&=& \left[2\tanh\left(-\frac{E_n-E_m}{2T_{B}}\right)\right] \bar{w}^{\beta}_{nm}
\ \ \ \ \ \ \
\\
p_n-p_m 
&=&  \left[2\tanh\left(-\frac{E_n-E_m}{2T_{nm}}\right)\right] \bar{p}_{nm}
\eeq 
At high temperatures one can approximate the $\tanh()$ 
by linear functions leading to 
\be{18}
\dot{\mathsf{Q}} &=& 
\frac{1}{2}\sum_{n,m} \bar{p}_{nm} \frac{\bar{w}^{\beta}_{nm}}{T_B} (E_n{-}E_m)^2
\\ \label{e19}
&-& 
\frac{1}{2}\sum_{n,m} \bar{p}_{nm} \frac{\bar{w}^{\beta}_{nm}}{T_{nm}} (E_n{-}E_m)^2
\eeq 
The first expression is identified as $D_B/T_B$, 
where $D_B$ is the bath induced diffusion coefficient.  
The second expression is used to define the effective 
system temperature~$ {T_{\tbox{sys}}}$, 
such that it takes the form $-D_B/ {T_{\tbox{sys}}}$. 

In the {\em stochastic} case the effect of the driving can be treated 
using the same procedure. The expression for $\dot{\mathsf{W}}$ 
is given by the first equality of \Eq{e2}, that is analogous to \Eq{e3}. 
Taking into account that $w_{nm}^{\varepsilon}$ unlike $w_{nm}^{\beta}$ 
is a symmetric matrix one obtains, in analogy to \Eq{e19}, 
\be{20}
\dot{\mathsf{W}} =
\frac{1}{2}\sum_{n,m} \bar{p}_{nm} \frac{w_{nm}^{\varepsilon}}{T_{nm}} (E_n{-}E_m)^2 
\eeq 
If the state is strictly canonical with all the $T_{nm}$ 
equal the same number $ {T_{\tbox{sys}}}$, 
then $\dot{\mathsf{W}} = D/ {T_{\tbox{sys}}}$ 
where $D$ is given by \Eq{e333}. This is the standard 
LRT expression for the diffusion coefficient, leading 
to the linear result~\Eq{e4} in the case of near-neighbor transitions.

More generally $ {T_{\tbox{sys}}}$~is the {\em effective}  
temperature, and the equation $\dot{\mathsf{W}}=D/ {T_{\tbox{sys}}}$ 
is used to {\em define} the effective diffusion coefficient~$D$, 
leading in the stochastic case to~\Eq{e243}.
This agrees with the linear result~\Eq{e4} for weak driving,  
and with the semi-linear result~\Eq{e5} for strong driving.


\noindent
{\bf Acknowledgments:}
This research has been supported by the US-Israel Binational Science Foundation (BSF).



\ \\

\clearpage

\newpage

\begin{figure}[h!t]
\centering

\includegraphics[width=0.9\hsize]{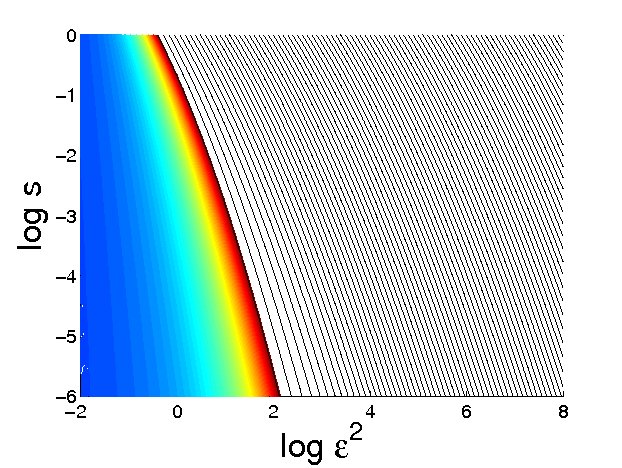}

\includegraphics[width=0.9\hsize]{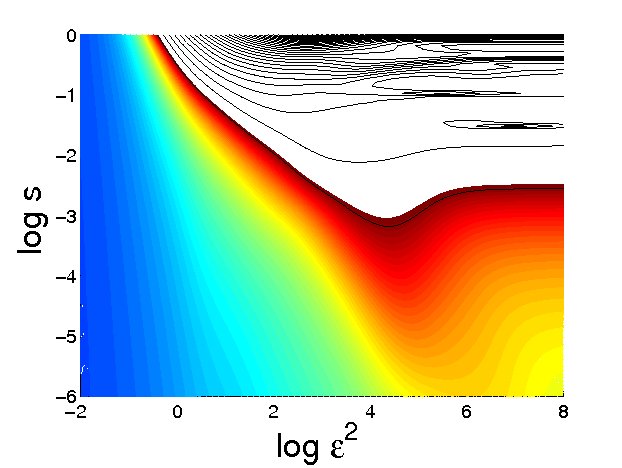}

\caption{
The effective NESS temperature $ {T_{\tbox{sys}}}$
in the stochastic case (upper panel) 
and in the quantum case (lower panel)     
is imaged for additional values of the sparsity.
 {(Color online) Blue represents the 
bath temperature ${T=10}$, 
while red corresponds to  ${T=50}$.
For quantitative dependence see \Fig{fT}.}
}

\label{fT2D}
\end{figure}

\clearpage

\begin{thebibliography}{99}
\bibitem[a]{a} 
There is an implicit conservation of probability requirement ${\sum p_n=1}$. 
The diagonal elements of $\mathcal{W}$ are $-\Gamma_n$, 
with the decay rates $\Gamma_n=\sum_m \mathcal{W}_{mn}$.
\bibitem[b]{b} 
Note that \Eq{e3} unlike \Eq{e242} can be trusted also in the quantum analysis, 
because the last term in \Eq{e10} does not couple diagonal and off-diagonal terms.
\bibitem[c]{c} 
The numerical demonstration in \Fig{TvsSigma} is done for pedagogical purpose,
to clarify the reasoning with regard to the dependence of $T_{\infty}$ on $\sigma$.
Obviously the numerics is dominated by the finite size truncation of energy space.
\end{thebibliography}
\end{document}